\def\Ar{\rightarrow}
\def\bar{\overline}

\def\a{\alpha}
\def\b{\beta}
\def\n{\nu}
\def\m{\mu}

\def\bar{\overline}

\def\eV{{\rm eV}}
\documentstyle [12pt,epsf]{article}
\newcommand{\lsim}{{\;\raise0.3ex\hbox{$<$\kern-0.75em\raise-1.1ex\hbox{$\sim$}}
\;}}
\newcommand{\gsim}{{\;\raise0.3ex\hbox{$>$\kern-0.75em\raise-1.1ex\hbox{$\sim$}}
\;}}
\begin{document}
\baselineskip=24.5 pt
\setcounter{page}{1}
\thispagestyle{empty}
%\topskip 2.5  cm
%\topskip 0.5  cm
%\begin{flushright}
%\begin{tabular}{l l}
%& {\normalsize hep-ph/99}\\
%& July 1999
%\end{tabular}
%\end{flushright}
\hspace{12.7cm} HIP-1999-54/TH\\

\vspace{2 cm}
\centerline{\Large\bf Search for CP Violation at a Neutrino Factory }

\centerline{\Large\bf in a Four-Neutrino Model }
\vskip 1.5 cm
\centerline{{ Anna Kalliom\"aki} $^a$
\footnote{E-mail address: amkallio@pcu.helsinki.fi},
 { Jukka  Maalampi} $^a$
 \footnote{E-mail address: maalampi@pcu.helsinki.fi}
 and  { Morimitsu Tanimoto} $^b$
  \footnote{E-mail address: tanimoto@edserv.ed.ehime-u.ac.jp}}
\vskip 0.8 cm
 \centerline{ \it{a) Theoretical Physics Division, Department of
Physics,
 University of Helsinki, Finland}}

 \centerline{ \it{b) Science Education Laboratory, Ehime University,
 Japan}}
\vskip 2 cm
\centerline{\bf ABSTRACT}\par
\vskip 0.5 cm

The CP violation effects in long baseline neutrino oscillations are
studied in the framework of a four-neutrino model (three active
neutrinos and one
sterile neutrino). It is assumed that neutrino masses are
devided into two nearly degenerate pairs, as indicated by the
oscillation data. Approximative analytic expressions are derived for the
probability differences $\Delta P_{\a\b} \equiv
 P(\nu_\alpha \rightarrow \nu_\beta)- P(\bar\nu_\alpha \rightarrow
\bar\nu_\beta)$ taking into account the CP violation effect of the
Earth's crust. The matter effect is found to be small compared with the
genuine CP violation term, in contrast with the three-neutrino model. 
\newpage
\topskip 0  cm
%%%%%%%%%%%%%%%%%%%%%%%%%%%%%%%%%%%%%%%%%%%%%%%%%%%%%%%%%%%%%%%%%%%%%%%%%%%%%%%
%%%%%%%%%%%%%%%%%%%%%%%%%%%%%%%%%%%%%%%%%%%%%%%%%%%%%%%%%%%%%%%%%%%%%%%%%%%%%%%

  A new stage of the neutrino oscillation studies is represented
by the long baseline (LBL)  oscillation  experiments.
 The LBL accelerator experiment K2K \cite{K2K} begins to take data
 in this year (1999), whereas the MINOS \cite{MINOS}
 and a CERN-to-Gran Sasso project \cite{ICARUS} will start in the first
year
 of the next millenium.
 Some authors \cite{MT,AS,MiNu,Bi,Barger} have  discussed  the
possibility of observing the leptonic CP violation  by measuring  in
LBL
experiments the difference of transition probabilities between
CP-conjugate neutrino oscillation
 channels \cite{CP0,CP}, such as $\n_\m \Ar\n_e$ and $\bar \n_\m \Ar
\bar \n_e$.
 The direct measurement is, however, very difficult in the planned LBL
 experiments since the magnitude of the probability difference $\Delta
P_{\a\b} \equiv
 P(\nu_\alpha \rightarrow \nu_\beta)- P(\bar\nu_\alpha \rightarrow
\bar\nu_\beta)$ is usually expected to be
 below $0.01$ and the difference of energy distributions of neutrino
beams $\n_\m$ and
 $\bar\n_\m$  disturbs this measurement at the level of  ${\cal O}(0.01)$.
 Moreover, the matter effects due to the Earth's crust, while not that
important for the oscillation probabilities themselves,  can be
sizeable for the probability differencies and make it difficult
to extract from the data the genuine CP violating effect caused by
complex elements of the neutrino mixing matrix. \par

 On the other hand, a {\it neutrino factory}, a high intensity and high
quality neutrino beam from a very intense muon source, has  recently
been under discussion in connection with the muon collider studies
\cite{Geer,Rujula}. It would provide an excellent possibility to search
for CP
 violation in  LBL neutrino oscillations because  the flavour content of
the beam would be known (50\% $\nu_{\mu}$ and 50\% $\overline\nu_{e}$ in
the case of negative muons), as would be also the  energy distributions
of the neutrinos as soon as the muon energy and polarization are given
\cite{Rujula}.

The magnitude of the CP violation in the neutrino oscillation depends
on
the number of neutrinos involved. One of the present authors (M.T.) has
studied prospects for searching the CP violation at a neutrino factory
in the scheme of three ordinary neutrinos $\nu_{e}$, $\nu_{\mu}$ and
$\nu_{\tau}$, and he proposed a method for extracting the  genuine CP
violation effect  from the observed signal \cite{cp3}.
It is not guaranteed, however, that the three left-handed active
neutrinos are the only neutrinos that exist in nature. Sterile
neutrinos, neutral leptons that lack the Standard Model interactions,
may also exist and they may give rise to interesting phenomena
\cite{sterile,four1,four2,BGG1,Barger1}. Actually, in order to explain
the
solar \cite{BKS},
atmospheric \cite{SKam} and LSND \cite{LSND} observations in terms of
neutrino oscillations one necessarily needs at least one sterile
neutrino in addition to the three ordinary neutrino flavours.

In this paper we shall study the CP violation effects in the
oscillations of four neutrinos and their signatures  in LBL oscillation
experiments at a neutrino factory. We assume that, besides the three
ordinary neutrinos $\nu_{e}$, $\nu_{\mu}$ and $\nu_{\tau}$, there exists
one sterile neutrino $\nu_{s}$. We shall make a comparison of the
predictions for the CP violation in the three-neutrino and
four-neutrino cases and discuss how these two cases could be
distinguished experimentally.

%%%%%%%%%%%%%%%%%%%%%%%%%%%%%%%%%%%%%%%%%%%%%%%%%%%%%%%%%%%%%%%%%%%%%%%%%%%%
 %%%  Four Neutrino Model  %%%

%%%%%%%%%%%%%%%%%%%%%%%%%%%%%%%%%%%%%%%%%%%%%%%%%%%%%%%%%%%%%%%%%%%%%%%%%%%%
 In order to be compatible with the solar \cite{BKS}, atmospheric
\cite{SKam}, LSND
data \cite{LSND}, as well as with the other accelerator and reactor
limits,
the neutrino masses in the four-neutrino model should be divided into
two  pairs of nearly degenerate masses \cite{BGG1,Barger1}, e.g.,
$\Delta m^2_{\odot} \equiv \Delta m^2_{01} \ll \Delta m^2_{atm} \equiv
\Delta m^2_{32} \ll \Delta m^2_{LSND} \equiv \Delta m^2_{21}$, where
$\Delta m_{ji}^2 = m_j^2-m_i^2$ and $m_i$ is the mass of the massive
neutrino state $\nu_i$ ($i=0,1,2,3$). One can assume without loss of
generality that $0 < m_0, m_1 < m_2 < m_3$.
If the solar neutrino oscillations are driven by the MSW effect
\cite{MSW},
one must require $m_0 > m_1$. In the following we will assume the
four-neutrino spectra to be like this.

 %%%%%%%%%%%%%%%%%%%%%%%%%%%%%%%%%%%%%%%%%%%%%%%%%%%%%%%%%%%%%
Neutrino oscillation probabilities $P(\nu_\alpha \rightarrow \nu_\beta)$
are determined by mass-squared  differences $\Delta m_{ji}^2$ and the
elements of the neutrino mixing matrix $U$ that connects the massive
neutrino states $\nu_i$ and the flavour neutrino states $\nu_{\alpha}$
through the relation $\nu_{\alpha}=\sum_iU_{\alpha i}\nu_i$ \cite{MNS}.
The
most
general mixing matrix $U$ for four Majorana neutrinos can be
parametrized in terms of
$6$ rotation angles and $6$ phases as follows:
\begin{equation}
U = \left( \begin{array}{cccc}
c_{01}c_{02}c_{03} & c_{02}c_{03}s_{01}^*
& c_{03}s_{02}^* & s_{03}^* \\
\\
-c_{01}c_{02}s_{03}s_{13}^*
& -c_{02}s_{01}^*s_{03}s_{13}^*
& -s_{02}^*s_{03}s_{13}^*
& c_{03}s_{13}^*
\\
-c_{01}c_{13}s_{02}s_{12}^*
& -c_{13}s_{01}^*s_{02}s_{12}^*
& +c_{02}c_{13}s_{12}^*
&
\\
-c_{12}c_{13}s_{01}
& +c_{01}c_{12}c_{13}
&
&
\\ \\
-c_{01}c_{02}c_{13}s_{03}s_{23}^*
& -c_{02}c_{13}s_{01}^*s_{03}s_{23}^*
& -c_{13}s_{02}^*s_{03}s_{23}^*
& c_{03}c_{13}s_{23}^*
\\
+c_{01}s_{02}s_{12}^*s_{13}s_{23}^*
& +s_{01}^*s_{02}s_{12}^*s_{13}s_{23}^*
& -c_{02}s_{12}^*s_{13}s_{23}^*
&
\\
-c_{01}c_{12}c_{23}s_{02}
& -c_{12}c_{23}s_{01}^*s_{02}
& +c_{02}c_{12}c_{23}
&
\\
+c_{12}s_{01}s_{13}s_{23}^*
& -c_{01}c_{12}s_{13}s_{23}^*
&
&
\\ +c_{23}s_{01}s_{12} &
-c_{01}c_{23}s_{12}
&
&
\\ \\
-c_{01}c_{02}c_{13}c_{23}s_{03}
& -c_{02}c_{13}c_{23}s_{01}^*s_{03}
& -c_{13}c_{23}s_{02}^*s_{03}
& c_{03}c_{13}c_{23}
\\
+c_{01}c_{23}s_{02}s_{12}^*s_{13}
& +c_{23}s_{01}^*s_{02}s_{12}^*s_{13}
& -c_{02}c_{23}s_{12}^*s_{13}
&
\\
+c_{01}c_{12}s_{02}s_{23}
& +c_{12}s_{01}^*s_{02}s_{23}
& -c_{02}c_{12}s_{23}
&
\\
+c_{12}c_{23}s_{01}s_{13}
& -c_{01}c_{12}c_{23}s_{13}
&
&
\\
-s_{01}s_{12}s_{23}
& +c_{01}s_{12}s_{23}
&
&
\\ \\ \end{array} \right) \ ,
\label{Mixing}
\end{equation}
\noindent
where $c_{ij}\equiv\cos\theta_{ij}$ and
$s_{ij}\equiv\sin\theta_{ij}e^{i\phi_{ij}}$ \cite{Notation}. If the mass
eigenstate
neutrinos are Dirac particles, only three phases are physically
meaningful.

The probability of the neutrino flavor oscillation
$\n_\a \Ar \n_\b$  ($\alpha,\beta=s,e,\mu,\tau$) in vacuum is given by
%%%%%%%%%%%%%%%%%%%%%%%%%%%%%%%%%
\begin{equation}
P(\nu_\alpha \rightarrow \nu_\beta) =
\delta_{\alpha\beta} - \sum_{j<k} \left[
  4 Y^{jk}_{\alpha\beta} \sin^2\Delta_{kj}
  - 2 J^{jk}_{\alpha\beta} \sin2\Delta_{kj} \right] \ ,
\label{prob}
\end{equation}
where
\begin{eqnarray}
Y^{jk}_{\alpha\beta} &\equiv
&
{\rm Re}(U_{\alpha j} U_{\alpha k}^* U_{\beta j}^* U_{\beta k}) \ ,
\qquad  J^{jk}_{\alpha\beta} \equiv {\rm Im}(U_{\alpha j} U_{\alpha k}^*
U_{\beta
j}^* U_{\beta k}) \ ,
\label{YJ}
\\
\Delta_{kj} &\equiv& \frac{\Delta m^2_{kj}}{4E}L \ , \qquad
\Delta m^2_{kj} \equiv m_k^2 - m_j^2 \ ,
\label{Delta}
\end{eqnarray}
$L$ is the oscillation distance, and $E$ is the neutrino energy.
 The   probability  $P(\bar\n_\a\Ar \bar\n_\b)$ of the CP conjugated
channel is obtained by
  replacing $U$ with $U^*$.

The difference of the transition probabilities between CP-conjugate
channels,
\begin{eqnarray}
\Delta P_{\a\b} &\equiv&
 P(\nu_\alpha \rightarrow \nu_\beta)- P(\bar\nu_\alpha \rightarrow
\bar\nu_\beta) \nonumber \\
 &=& 4  ( J^{01}_{\alpha\beta} \sin2\Delta_{10} + J^{02}_{\alpha\beta}
\sin2\Delta_{20}+
 J^{03}_{\alpha\beta} \sin2\Delta_{30} \nonumber\\
 &+& J^{12}_{\alpha\beta} \sin2\Delta_{21} + J^{13}_{\alpha\beta}
\sin2\Delta_{31}+
     J^{23}_{\alpha\beta} \sin2\Delta_{32})  \ ,
\label{CP1}
\end{eqnarray}
directly measures  the CP violation that originates from the complex
phases of the mixing matrix $U$   \cite{CP0,CP}.
Basically due to unitarity of the mixing matrix $U$, only three of the
differences $\Delta P_{\alpha\beta}$ are independent in the
four-neutrino case, and, correspondingly, only three out of the six
phases of the mixing matrix $U$ of Majorana neutrinos can be determined
by neutrino oscillation measurements. Therefore, as far as neutrino
oscillations are concerned, our considerations apply to both Dirac and
Majorana neutrinos. The three
remaining independent phases of $U$ enter into the mass matrix
elements and appear in lepton number violating processes involving
Majorana neutrinos, such as neutrinoless double beta decay.

By using the relations
\begin{equation}
J^{12}_{\alpha\beta} = J^{23}_{\alpha\beta}-J^{02}_{\alpha\beta} \ ,
\qquad
J^{13}_{\alpha\beta} = -J^{23}_{\alpha\beta}-J^{03}_{\alpha\beta} \ ,
\end{equation}
one can rewrite (\ref{CP1}) in the form
\begin{eqnarray}
\Delta P_{\a\b} &=&
 - 4  J^{23}_{\alpha\beta}
(\sin2\Delta_{12}+\sin2\Delta_{23}+\sin2\Delta_{31})
  + 4   J^{01}_{\alpha\beta} \sin2\Delta_{10}     \nonumber \\
  & + & 4 J^{02}_{\alpha\beta} (\sin2\Delta_{20}- \sin2\Delta_{21})
 + 4 J^{03}_{\alpha\beta} (\sin2\Delta_{30}- \sin2\Delta_{31}).
\label{CP2}
\end{eqnarray}
Since
\begin{eqnarray}
\sin2\Delta_{20}- \sin2\Delta_{21}&=&2
\cos(\Delta_{20}+\Delta_{21})\sin\Delta_{10} \ , \nonumber \\
\sin2\Delta_{30}- \sin2\Delta_{31}&=&2
\cos(\Delta_{30}+\Delta_{31})\sin\Delta_{10} \ ,
\end{eqnarray}
one can see that the last two terms in eq.~(\ref{CP2}) average out to
zero in the LBL experiments due to the rapidly oscillating cosines.

%%%%%%%%%%%%%%%%%%%%%%%%%%%%%%%%%%%%%%%%%%%%%%%%%%%%%%%%%%%%%%%%%%%%%

 In the neutrino factory one has a $\n_\m + \bar\n_e\ (\bar\n_\m +
\n_e)$ beam originating
 in the decay of $\m^- \ (\m^+)$. As mentioned above, there are only
three independent probability differences
 $\Delta P_{\a\b}$ from which information about the CP violation can
be inferred.
 Let us consider the probability differences associated with the
channels $\n_e\Ar\n_\mu$, $\n_\mu\Ar\n_e$, and $\n_e\Ar\n_\tau$, which are observable in practice. They
are given as
\begin{eqnarray}
\Delta P_{e\mu} &=&
 - 4 J^{23}_{e\mu} (\sin2\Delta_{12}+\sin2\Delta_{23}+\sin2\Delta_{31})
 + 4 J^{01}_{e\mu} \sin2\Delta_{10} \ ,\nonumber \\
\Delta P_{\mu e} &=&
 - 4 J^{23}_{\mu e} (\sin2\Delta_{12}+\sin2\Delta_{23}+\sin2\Delta_{31})
 + 4 J^{01}_{\mu e} \sin2\Delta_{10} \ ,\nonumber \\
 \Delta P_{e\tau} &=&
 - 4 J^{23}_{e\tau} (\sin2\Delta_{12}+\sin2\Delta_{23}+\sin2\Delta_{31})
 + 4 J^{01}_{e\tau} \sin2\Delta_{10} \  .
 \label{CP3}
\end{eqnarray}

In order to be able to estimate $J^{01}_{\a\b}$, we need information of
mixings between the sterile neutrino $\n_s$ and left-handed active
neutrinos $\nu_{e}$, $\nu_{\mu}$ and $\nu_{\tau}$. These mixings are
severely constrained by the standard Big Bang cosmology.
If the mixing is too large, neutrino oscillations, acting as an
effective interaction, would bring the sterile neutrino in equilibrium
before neutrino decoupling, and the resulting excess in energy density
would endanger the standard scheme for the nucleosynthesis of light
elements (BBN) \cite{cosc}. Moreover, a new mechanism of resonant
sterile neutrino conversion in the boundaries of spatial opposite-sign
lepton number domains was recently proposed \cite{ShFu} and was found to
lead  to even more stringent constraints on the mixings, in particular
for large mass-squared differencies. The ensuing constraints are
summarized as follows \cite{ShFu}:
for the  $\n_{\mu,\tau}-\n_{s}$ mixings,
\begin{eqnarray}
  |\Delta m^2| \sin^2 2\theta < 7\times 10^{-5}\,\eV^2 \ {\rm for} \
                                       |\Delta m^2|\leq 2.5\times
10^3\,\eV^2 \ , \nonumber\\
  \sin^2 2\theta < 3\times 10^{-8}\ {\rm for}\
                                       |\Delta m^2|\geq 2.5\times
10^3\,\eV^2 \ ,
\end{eqnarray}
\noindent and for the $\n_{e}-\n_{s}$ mixing,
\begin{eqnarray}
  |\Delta m^2| \sin^2 2\theta < 5\times 10^{-8}\,\eV^2\  {\rm for} \
                                       |\Delta m^2|\leq 4\,\eV^2 \ ,
\nonumber\\
  \sin^2 2\theta < 10^{-8}\ {\rm for}\
                                       |\Delta m^2|\geq 4\,\eV^2 .
\label{Shilimit}
\end{eqnarray}
 From this we conclude that the $\n_{\mu,\tau}-\n_{s}$ mixings can be
neglected in the following study as the corresponding
mass-squared differences are supposed to be of the order of $\Delta
m_{\rm LSND}^2\simeq 1 \: {\rm eV}^2$.
 On the other hand, for the $\n_{e}-\n_{s}$ mixing, which is assumed to
be responsible for the solar neutrino deficit, the mass-squared
difference is very small, so that (\ref{Shilimit}) does not constrain
that mixing severely.

 Neglecting the $\n_{\mu,\tau}-\n_{s}$ mixings, the matrix $U$ is
expressed as follows:
%%%%%%%%%%%%%%%%%%%%%%%%%%%%%%%%%%%%%%%%%%%%%%%%%%%%%%%%%%%%
\begin{equation}
U \simeq \left( \begin{array}{cccc}
c_{01} & s_{01}^* & s_{02}^* & s_{03}^*
\\
&&&\\
-s_{01} & c_{01} & s_{12}^* & s_{13}^*
\\
&&&\\
-c_{01}(s_{23}^*s_{03}+c_{23}s_{02})
& -s_{01}^*(s_{23}^*s_{03}+c_{23}s_{02})
& c_{23}
& s_{23}^*
\\
+s_{01}(s_{23}^*s_{13}+c_{23}s_{12})
& -c_{01}(s_{23}^*s_{13}+c_{23}s_{12})
&&\\
&&&\\
c_{01}(s_{23}s_{02}-c_{23}s_{03})
& s_{01}^*(s_{23}s_{02}-c_{23}s_{03})
& -s_{23}
& c_{23}
\\
-s_{01}(s_{23}s_{12}-c_{23}s_{13})
& +c_{01}(s_{23}s_{12}-c_{23}s_{13})
&&\\
&&&\\
\end{array} \right) \ .
\label{AppU}
\end{equation}
%%%%%%%%%%%%%%%%%%%%%%%%%%%%%%%%%%%%%%%%%%%%%%%%%%%%%%%
By using this approximation, we have
\begin{equation}
J^{23}_{e\mu} = - J^{23}_{e\tau} = -J^{23}_{\mu e}=
 c_{23}s_{12}s_{23}s_{13}\sin \phi\equiv J_{CP} \ ,
\end{equation}
\noindent
where $\phi=\phi_{13}-\phi_{12}-\phi_{23}$.
$J_{CP}$ is the four-neutrino counterpart of the famous Jarlskog
invariant \cite{Cecilia} defined for the three flavour mixing.
Then, eqs.~(\ref{CP3}) become
\begin{eqnarray}
\Delta P_{e\mu} &=&
 - 4 J_{CP} (\sin2\Delta_{12}+\sin2\Delta_{23}+\sin2\Delta_{31})
 + 4 J^{01}_{e\mu} \sin2\Delta_{10} \ ,\nonumber \\
\Delta P_{\mu e} &=&
  4 J_{CP} (\sin2\Delta_{12}+\sin2\Delta_{23}+\sin2\Delta_{31})
 + 4 J^{01}_{\mu e} \sin2\Delta_{10} \ ,\nonumber \\
 \Delta P_{e\tau} &=&
  4 J_{CP} (\sin2\Delta_{12}+\sin2\Delta_{23}+\sin2\Delta_{31})
 + 4 J^{01}_{e\tau} \sin2\Delta_{10} \  .
 \label{CP4}
\end{eqnarray}
\noindent
 Taking account of $J^{01}_{e\mu}=-J^{01}_{\mu e}$, we get the CPT invariant relation $\Delta P_{e\mu} = - \Delta P_{\mu e}$, which is the same relation as in the three family model.

 How large is $J^{01}_{e\mu}$?
 In  the approximate mixing matrix $U$, that is when $s_{02}=0$ and
$s_{03}=0$,
 we have $J^{01}_{e\mu}=J^{01}_{e\tau}=0$.
 Therefore, the CP violating effect follows from the active neutrino
mixings and phases. The mixing between the sterile and the active
neutrinos
 contributes to $\Delta P_{\a\b}$ only indirectly through the unitarity
of
the $4\times 4$
  mixing matrix.

 %%%%%%%%%%%%%%%%%%%%%%%%%%%%%%%%%%%%%%%%%%%%%%%%%%%%%%%%%%%
 %%%%%%%%%%%%%%%%%%%  Matter Effect  %%%%%%%%%%%%%%%%%%%%%%%
 %%%%%%%%%%%%%%%%%%%%%%%%%%%%%%%%%%%%%%%%%%%%%%%%%%%%%%%%%%%

The above-derived CP violation measures $\Delta P_{\a\b}$ as such should
not  be compared with observations but one should take into account also
matter effects in the earth. The earth is CP-odd in the sense that it
acts differently in the propagation of neutrinos and antineutrinos. It
will therefore affect $\Delta P_{\a\b}$, even if the distance travelled
by neutrinos is less than $1000\, {\rm Km}$, like it is in LBL
experiments. The matter effect should be carefully analyzed since it
depends strongly on the mass hierarchy, mixings and the incident energy
of the neutrino, as was shown in the previous works
\cite{matter,matter1,matter2}. The effective mass squared in the matter,
$M_m^2$, for the neutrino energy $E$ is in weak basis given by

 %%%%%%%%%%%%%%%%%%%%%%%%%
\begin{equation}
        M_m^2 = U \left (\matrix{0 & 0 & 0 & 0 \cr
            0 & \Delta m_{10}^2 & 0     & 0 \cr
            0 & 0     & \Delta m_{20}^2 & 0 \cr
            0 & 0 & 0 & \Delta m_{30}^2     \cr} \right )U^\dagger +
      \left (\matrix{ 2 E a' & 0 & 0 & 0 \cr 0 & 2 E a & 0 & 0 \cr
            0 & 0 & 0 & 0 \cr          0 & 0 & 0 & 0 \cr} \right )  \ ,
			\label{mass}
\end{equation}
\noindent
where $a= \sqrt{2} G_F n_e$ and $a'= \sqrt{2} G_F n_e/2$.
 For antineutrinos, the effective mass squared is obtained by
 replacing $a\Ar -a$, $a'\Ar -a'$ and $U \Ar U^*$.

 Since $\Delta m_{10}^2 \ll a, a'$ in  LBL experiments,
 the sterile neutrino decouples from
 active ones as far as $s_{02}\simeq s_{03}\simeq 0$.
 Therefore it is enough to discuss the matter effect on the active
neutrinos only in LBL experiments
\footnote{In short baseline experiments and the solar neutrino
experiment,
the sterile neutrino couples with active ones as far as $s_{01}\not =
0$.}.
 %%%%%%%%%%%%%%%%%%%%%%%%%%%%%%%%%%%%%%%%%%%%%%%%%%%%%%%%%%%%%%%%%%%%%
 %%%%%%%%%%%%%%%%%%%%%%%%%%%%%%%%%%%%%%%%%%%%%%%%%%%%%%%%%%%%%%%%%%%%%%

  Although we can calculate  $\Delta P_{\a\b}$ in the presence of matter
numerically,
  an approximate analytic formula
  is useful for investigating the qualitative structure of the matter
effect.
  For the case that the highest squared-mass difference is ${\cal O}(1
\eV^2)$
    the appropriate formulae have been given in the lowest order
  approximation  by Minakata and Nunokawa \cite{MiNu}:

 \begin{eqnarray}
\Delta P_{\a\b}=
 &-& 4 \sum_{j>i} \mbox{Re}(UUUU)_{\alpha\beta \: ;\:  ij}
\biggl[\sin^2\{\frac{1}{2}I_{ij}(a)\}-\sin^2\{\frac{1}{2}I_{ij}(-a)\}\biggr]
\cr
&+&2 \sum_{j>i}\mbox{Im}(UUUU)_{\alpha\beta \: ;\: ij}
\biggl[\sin{I_{ij}(a)} + \sin{I_{ij}(-a)}\biggr] \cr
&-& 4 \sum_{j>i} \mbox{Re}(UUU \delta V)_{\alpha\beta \: ;\:  ij}
\biggl[\sin^2\{\frac{1}{2}I_{ij}(a)\}+\sin^2\{\frac{1}{2}I_{ij}(-a)\}\biggr]
\cr
&+& 2 \sum_{j>i}\mbox{Im}(UUU\delta V)_{\alpha\beta \: ;\: ij}
\biggl[\sin{I_{ij}(a)} - \sin{I_{ij}(-a)}\biggr] \ ,
\label{matt1}
\end{eqnarray}
where
\begin{equation}
(UUUU)_{\alpha\beta \: ;\: ij}
= U_{\alpha i}^*U_{\alpha j}U_{\beta i}U^*_{\beta j}
\end{equation}
\noindent and
$\delta V$ denotes the first order correction for the mixing matrix in
vacuum as
\begin{equation}
 \delta V_{\a i} =2Ea\sum_{j\not= i} \frac{U_{\a
j}U^*_{ej}U_{ei}}{\Delta
m^2_{ij}} \ .
 \label{dV}
\end{equation}
The last term in (\ref{matt1}) is at least second order in $a$.
%%%%%%%%%%%%%%%%%%%%%%%%%%%%%%%%%%%%%%%%%
The quantity $I_{ij}(a)$ appearing in (\ref{matt1}) is given by
\begin{equation}
I_{ij}(a)=\frac{\Delta m^2_{ij}}{2E}L + (|U_{ei}|^2-|U_{ej}|^2)
\int_0^L {\rm d}x a(x) \ .
\end{equation}
Eq.~(\ref{matt1}) relies only on the hierarchy
$a \ll |\Delta m^2_{ij}| /E$ but is not affected by any possible
hierarchy among $\Delta m^2_{ij}$.

%%%%%%%%%%%%%%%%%%%%%%%%%%%%%%%%%%%%%%%%%%%%%%%%%%%%%%%%%%%%%%%%%%%%%%
We now approximate the density of the Earth by a constant  ($\rho=2.8\,
{\rm g}/{\rm cm}^3$), i.e. we take $a$ as a constant~\footnote{
 The constant matter density is not always a particularly good
approximation quantitatively, so that for an accurate estimation of the
matter effect one should apply a real earth model with an appropriately
varying density. Actually, Koike and
Sato have discussed  the matter effect in the K2K experiment by using a
real earth model \cite{KoSa}.}.
%%%%%%%%%%%%%%%%%%%%%%%%%%%%%%%%%%%%%%%%%%%%%%%%%%%%%%%%%%%%%%%%%%%%%%%
 By using eqs.~(\ref{matt1}) and  (\ref{dV}),  we get
\footnote{The case of $\Delta m^2_{31}\simeq \Delta m^2_{32} \gg \Delta
m^2_{12}$ was
 discussed in ref.\cite{MiNu}.}
\begin{eqnarray}
\Delta P_{\a\b}
&& =
 \pm 4 J_{CP} \cos\biggl[\biggl(|U_{e2}|^2- |U_{e3}|^2\biggr)aL\biggr]
\sin\frac{\Delta m^2}{2E} L \cr
&&
+ 4 \mbox{Re}(U_{\b 2}U_{\b 3}^*U_{\a 2}^*U_{\a 3})
\sin\biggl[\biggl(|U_{e2}|^2- |U_{e3}|^2\biggr) aL\biggr]
\sin\frac{\Delta m^2}{2E} L \cr
&& - \frac{16Ea}{\Delta m^2}
\mbox{Re}\biggl[U_{\b 2}U_{\b 3}^* U^*_{e 2}U_{e 3}
\biggl(|U_{\a 2}|^2- |U_{\a 3}|^2 \biggr)  \cr
&& + U_{\a 2}^* U_{\a 3} U_{e 2}U^*_{e 3}
  \biggl(|U_{\b 2}|^2- |U_{\b 3}|^2 \biggr) \biggr]
\sin^2\frac{\Delta m^2}{4E} L \cr
&& -\frac{16Ea}{\Delta M^2}
\mbox{Re}\biggl[U_{\b 2}U^*_{\b 3}
(U^*_{\a 2}U_{\a 1}U_{e3}U^*_{e1}
+ U_{\a 3}U^*_{\a 1}U^*_{e2}U_{e1}) \cr
&& + U^*_{\a 2}U_{\a 3}(U_{\b 2}U^*_{\b 1}U^*_{e3}U_{e1}
+ U^*_{\b 3}U_{\b 1}U_{e2}U^*_{e1}) \biggr]
\sin^2\frac{\Delta m^2}{4E}L \cr
&& +\frac{32Ea}{\Delta M^2}
|U_{e 1}|^2 \biggl[
2|U_{\a 1}|^2|U_{\b 1}|^2 - \delta_{\alpha e}|U_{\beta 1}|^2
- \delta_{\beta e}|U_{\alpha 1}|^2 \biggr]
\sin^2\frac{\Delta M^2}{4E}L \ ,
\label{Mappro0}
\end{eqnarray}
where the last term of eq.~(\ref{matt1}) proportional to
$\mbox{Im}(UUU\delta V)$ has been ignored because it is of order $a^2$
or higher.
The $\Delta M^2$ and $\Delta m^2$ denote
 $\Delta m_{31}^2\simeq \Delta m_{21}^2\simeq {\cal O}(\eV^2)$ and
 $\Delta m_{32}^2\simeq 10^{-3}\eV^2$, respectively. Since $\Delta
M^2\gg \Delta m^2,$ the terms proportional to $a/\Delta M^2$ can be
safely neglected in (\ref{Mappro0}). The first term in (\ref{Mappro0})
is the genuine CP violation and the second and third terms are leading
matter corrections.
We can write approximate formulae for the $\n_e\Ar \n_\mu$, $\n_e\Ar
\n_\tau$
and  $\n_\mu\Ar \n_e$ channels in terms of the mixing parameters as
follows:
%%%%%%%%%%%%%%%%%%%%%%%%%%%%%%%%%%%%%%%%%%%%%
\begin{eqnarray}
\Delta P_{e\mu}  &=&
 4 J_{CP} \sin\frac{\Delta m^2}{2E} L \cr
&&
+ 4 aL(J_{CP} \cot\phi - c_{13}^2 s_{13}^2 s_{12}^2s_{23}^2)
 (c_{13}^2 s_{12}^2- s_{13}^2)
\sin\frac{\Delta m^2}{2E} L\cr
&& - \frac{16Ea}{\Delta m^2}
\biggl [(J_{CP} \cot\phi - c_{13}^2 s_{13}^2 s_{12}^2s_{23}^2)(c_{13}^2
s_{12}^2- s_{13}^2)\cr
 && -s_{12}^2s_{13}^2 \biggl \{c_{13}^2 s_{23}^2
(c_{12}^2+c_{13}^2-s_{12}^2s_{13}^2)
 -c_{12}^2c_{13}^2+ 2 J_{CP}\cot\phi \biggr \}  \biggr ]
\sin^2\frac{\Delta m^2}{4E} L \ ,
\label{Mappro1}
\end{eqnarray}
%%%%%%%%%%%%%%%%%%%%%%%%%%%%%%%%%%%%%%%%%%%%%%
\begin{eqnarray}
\Delta P_{e\tau} &=&
 -4 J_{CP}\sin\frac{\Delta m^2}{2E} L \cr
&&
- 4 aL(J_{CP}\cot\phi + c_{13}^2 s_{13}^2 s_{12}^2 c_{23}^2)
 (c_{13}^2 s_{12}^2- s_{13}^2)
\sin\frac{\Delta m^2}{2E} L \cr
&& + \frac{16Ea}{\Delta m^2}
 \biggl [(J_{CP}\cot\phi + c_{13}^2 s_{13}^2 s_{12}^2 c_{23}^2)(c_{13}^2
s_{12}^2- s_{13}^2)\cr
 && +s_{12}^2 s_{13}^2 \biggl \{c_{13}^2 c_{23}^2
(c_{12}^2+c_{13}^2-s_{12}^2s_{13}^2)
 -c_{12}^2c_{13}^2 - 2 J_{CP}\cot\phi \biggr \}   \biggr ]
\sin^2\frac{\Delta m^2}{4E} L \ ,
\label{Mappro2}
\end{eqnarray}
%%%%%%%%%%%%%%%%%%%%%%%%%%%%%%%%%%%%%%%%%%%%%%
\begin{eqnarray}
\Delta P_{\mu e} &=&
 -4 J_{CP} \sin\frac{\Delta m^2}{2E} L \cr
&&
+ 4 aL(J_{CP} \cot\phi - c_{13}^2 s_{13}^2 s_{12}^2s_{23}^2)
 (c_{13}^2 s_{12}^2- s_{13}^2)
\sin\frac{\Delta m^2}{2E} L \cr
&& - \frac{16Ea}{\Delta m^2}
 \biggl [(J_{CP} \cot\phi - c_{13}^2 s_{13}^2 s_{12}^2s_{23}^2)(c_{13}^2
s_{12}^2- s_{13}^2) \cr
 && -s_{12}^2s_{13}^2 \biggl \{c_{13}^2 s_{23}^2
(c_{12}^2+c_{13}^2-s_{12}^2s_{13}^2)
 -c_{12}^2c_{13}^2+ 2 J_{CP} \cot\phi \biggr \} \biggr ]
\sin^2\frac{\Delta m^2}{4E} L \ .
\label{Mappro3}
\end{eqnarray}
%%%%%%%%%%%%%%%%%%%%%%%%%%%%%%%%%%%%%%%%%%%%%%%
Taking account of  $s_{12}, s_{13} \ll 1$,
these formulae are expressed as
 \begin{eqnarray}
\Delta P_{e\mu} &\simeq&  4 J_{CP} \sin\frac{\Delta m^2}{2E} L
               +( s_{23}^2- \frac{1}{2}) K_{matt} + J_{matt} \ ,
\nonumber \\
\Delta P_{e\tau} &\simeq&  -4 J_{CP} \sin\frac{\Delta m^2}{2E} L
               +( c_{23}^2- \frac{1}{2}) K_{matt} - J_{matt} \ ,
\nonumber \\
\Delta P_{\mu e} &\simeq&  -4 J_{CP} \sin\frac{\Delta m^2}{2E} L
+( s_{23}^2- \frac{1}{2}) K_{matt} + J_{matt}  \ ,
\label{Mappro4}
\end{eqnarray}
\noindent where
\begin{eqnarray}
K_{matt}&\simeq& \frac{32Ea}{\Delta m^2} s_{12}^2 s_{13}^2
               \sin^2\frac{\Delta m^2}{4E} L \ ,  \nonumber \\
J_{matt}&\simeq& 4 J_{CP}\cot\phi (s_{12}^2 - s_{13}^2)
               \biggl (aL\sin \frac{\Delta m^2}{2E} L -
               \frac{4aE}{\Delta m^2} \sin^2 \frac{\Delta m^2}{4E} L
\biggr ) \ .
\label{KJ}
\end{eqnarray}
 %%%%%%%%%%%%%%%%%%%%%%%%%%%%%%%%%%%%%%%%%%%%%%%%%%%
 %%%% Numerical Discussion
 %%%%%%%%%%%%%%%%%%%%%%%%%%%%%%%%%%%%%%%%%%%%%%%%%%%

We have checked that the approximate results in  eq.~(\ref{Mappro4}) are
in agreement with numerically computed results within $1 \%$
for the parameter values $L=732\,{\rm km}$ and  $E=7\,{\rm GeV}$.
According to eq.~(\ref{KJ}) the matter effect is of the order of
$s^2_{12}s^2_{13}$.
 The mixing parameters $s_{12}$ and $s_{13}$ are constrained by the
accelerator and reactor data. An upper limit $s_{12}^2 s_{13}^2 \lsim
10^{-4}$ can be inferred from the results of the Bugey disappearance
experiment \cite{Dec}. By inserting $s_{23}\simeq 1/\sqrt{2}$, as
indicated by the atmospheric neutrino data, one obtains $|J_{CP}|\lsim
0.005\sin\phi\,$ and thus, by using the relevant parameter values
$L=732\,
{\rm km}$,
 $E=7\,{\rm GeV}$ and $\rho=2.8\,{\rm g}/{\rm cm}^3$,
 one has $|K_{matt}|\lsim 4\times 10^{-5}$ and $|J_{matt}|\lsim
10^{-5}$.
 On the other hand, the genuine CP violation $4 J_{CP}
\sin (\Delta m^2L/2E)$ could be as large as ${\cal O}(10^{-2})$.
 Thus the matter effect is quite tiny in the four-neutrino model, in
contrast with the three-neutrino model where it may with certain
parameter values be as important as the genuine CP violation term
\cite{cp3}.

 From (\ref{Mappro4}) one obtains the relations
%%%%%%%%%%%%%%%%%%%%%%%%%%%%%%%%%%%%%%%%%%%%%%
\begin{equation}
 \Delta P_{e\mu}+\Delta P_{e\tau}=0 \ ,
 \qquad\quad
 \Delta P_{e\mu} - \Delta P_{\mu e}= 8 J_{CP}  \sin^2\frac{\Delta
m^2}{4E} L  \ .
\end{equation}
\noindent
The first relation shows that in the four-neutrino model the vacuum
relation $\Delta P_{e\mu}=-\Delta P_{e\tau}$ is not spoiled by matter
effect, in contrast with the three-neutrino case where in general
$|\Delta P_{e\tau}|>|\Delta P_{e\mu}|$ and for some part of the allowed
parameter space even $|\Delta P_{e\tau}|\gg |\Delta P_{e\mu}|$ due to
the matter effects (if we assume the sine of the phase angle to be positive;
in the opposite case the relations turn around \cite{cp3}).
 The matter independence of the second relation is guaranteed by CPT
invariance.
In the three-neutrino model \cite{cp3} one has
%%%%%%%%%%%%%%%%%%%%%%%%%%%%%%%%%%%
 %%%%%%%%%%%%  3 family  %%%%%%%%%%%%%
 %%%%%%%%%%%%%%%%%%%%%%%%%%%%%%%%%%%
\begin{eqnarray}
 \Delta P_{e\mu} &\simeq& 4 J_{CP}' f_{CP} + P_m
s^2_{23} \ , \nonumber\\
 \Delta P_{e\tau}&\simeq& -4 J_{CP}' f_{CP} + P_m
c^2_{23} \ , \nonumber\\
 \Delta P_{\mu e}&\simeq& -4 J_{CP}' f_{CP} + P_m
s^2_{23} \ ,
 \label{3CP}
 \end{eqnarray}
 \noindent   where $J_{CP}'$ is the Jarlskog parameter in the
three-neutrino case,
 \begin{eqnarray}
 P_m &\simeq & 8 s_{13}^2 \left (
 \frac{4 aE}{\Delta m_{31}^2}\sin^2 \frac{\Delta m_{31}^2L}{4E} -
 \frac{aL}{2}\sin \frac{\Delta m_{31}^2L}{2E} \right ) \ , \nonumber\\
f_{CP} &\equiv & \sin\frac{\Delta m^2_{12} L}{2E}+\sin\frac{\Delta
m^2_{23} L}{2E}+\sin\frac{\Delta m^2_{31} L}{2E} \ ,
 \label{f3}
 \end{eqnarray}
 \noindent  and two independent mass differences are $-\Delta
m^2_{12}\equiv \Delta m^2_{\odot}\simeq 10^{-5} \sim
10^{-4}\,\eV^2 $ and $\Delta m_{32}^2\equiv \Delta m^2_{atm}
\simeq 10^{-3}\,\eV^2$. In formulas (\ref{3CP}) the value of $J_{CP}'$,
unlike $J_{CP}$ in the four-neutrino case, depends on the solution of
the solar neutrino problem.
 %%%%%%%%%%%%%%%%%%%%%%%%%%%%%%%%%%%%%%%%%%%%%%%%%%%
 Because of the differences in equations (\ref{Mappro4}) and
(\ref{3CP}), it may be possible to distinguish the three-neutrino model
from the four-neutrino model by measuring the quantity $\Delta P_{\a\b}
$ in all three channels.

 %%%%%%%%%%%%%%%%  Conclusion  %%%%%%%%%%%%%%%%%%%%%
 %%%%%%%%%%%%%%%%%%%%%%%%%%%%%%%%%%%%%%%%%%%%%%%%%%%

In summary, we have studied the CP violation in four-neutrino
oscillations in a long baseline at a neutrino factory. We have evaluated
the probability differences $\Delta P_{\alpha\beta}\equiv
P(\nu_{\alpha}-\nu_{\beta})-P(\bar{\nu}_{\alpha}-\bar{\nu}_{\beta})$
between CP-conjugate channels and compared them with the corresponding
results in the three-neutrino case.  For the analytical expressions of
these quantities we have used an approximation in which the mixing of
the sterile neutrino with the muon and tau neutrinos are neglected. This
approximation is justified by cosmological constraints on the
active-sterile mixings. The  matter effect on the CP violation has
been derived following the approximative approach of Minakata and
Nunokawa~\cite{MiNu}. We conclude that in the four-neutrino model the
matter effect is generally small compared with the genuine CP
violation term, which is not so for the three-neutrino case. We also
show that probability differences $\Delta P_{\alpha\beta}$ between
CP-conjugate channels have different relative magnitudes in the
three-neutrino and four-neutrino models. In future, this may provide a
way to distinguish between these two models.

 %\vskip 1 cm
\section*{Acknowledgments}
This work was supported by the Academy of Finland under the project no. 40677 and 
by the Grant-in-Aid for Science Research, Ministry of Education, Science and Culture, Japan(No.10640274 ).
%%%%%%%%%%%%%%%%%%%%%%%%%%%%%%%%%%%%%%%%%%%%%%%%%%%%%%%%%%%%%%%%%
%%%%%%%%%%%%%%%%%%%%%%%%%%%%%%%%%%%%%%%%%%%%%%%%%%%%%%%%%%%%%%%%%
%\newpage
%%%%%%%%%%

\end{document}